\documentclass[12pt]{iopart}

\usepackage{graphicx}
\usepackage{epsfig}
\bibstyle{apsrev.bib}

\begin{document}
\input epsf.sty

\title[Entanglement entropy in aperiodic singlet phases]{Entanglement entropy in aperiodic singlet phases}

\author{R\'obert Juh\'asz$^1$ and Zolt\'an Zimbor\'as$^2$}

\address{$^1$ Research Institute for Solid
State Physics and Optics, H-1525 Budapest, P.O.Box 49, Hungary}
\ead{juhasz@szfki.hu}
\address{$^2$ Research Institute for Particle and Nuclear Physics, 
H-1525 Budapest, P.O.Box 49, Hungary}
\ead{zimboras@rmki.kfki.hu}

\begin{abstract}
We study the average entanglement entropy of blocks of contiguous spins in
aperiodic XXZ chains which possess an aperiodic singlet phase at
least in a certain limit of the coupling ratios. 
In this phase, where the ground state constructed
by a real space renormalization group method, consists 
(asymptotically) of independent singlet pairs, 
the average entanglement entropy is found to be a piecewise linear
function of the block size. The enveloping curve of this function is
growing logarithmically with the block size, with an effective
central charge in front of the logarithm which is characteristic 
for the underlying aperiodic sequence. 
The aperiodic sequence producing the largest effective central
charge is identified, and the latter is found to exceed the central 
charge of the corresponding homogeneous model.   
For marginal aperiodic modulations, 
numerical investigations performed for the XX model show a logarithmic
dependence, as well, with an effective central charge varying continuously 
with the coupling ratio.
\end{abstract}
\submitto{Journal of Statistical Mechanics: Theory and Experiment}
\maketitle

\newcommand{\bc}{\begin{center}}
\newcommand{\ec}{\end{center}}
\newcommand{\be}{\begin{equation}}
\newcommand{\ee}{\end{equation}}
\newcommand{\beqn}{\begin{eqnarray}}
\newcommand{\eeqn}{\end{eqnarray}}

\vskip 2cm

\section{Introduction}
The entanglement properties of strongly correlated quantum systems
is presently attracting great attention in
condensed matter physics \cite{VLRK,CC,RM,fazio}, 
quantum information theory \cite{QI},
and DMRG theory \cite{DMRG}.
From the point of view of condensed matter physics,
one main aim is to understand the role of quantum correlations 
in quantum critical phenomena.
A widely used estimator of quantum correlations is the
{\it entanglement entropy} of a subsystem $s$ embedded in a larger system, 
which is defined as
\begin{equation}
S(\hat{\rho}_{s})=-{\rm Tr} \; \hat{\rho}_{s} \ln \hat{\rho}_{s} \; , \nonumber
\end{equation} 
where $\hat{\rho}_{s}$ denotes the reduced density matrix of the
subsystem. Near quantum critical points entanglement entropy was 
found to display interesting scaling behavior
\cite{VLRK,CC,Qfree,Ddim}.
In one dimensional homogeneous critical spin chains the entanglement 
entropy belonging to a block of $L$ contiguous spins was shown to grow 
asymptotically as 
\begin{equation}
S_L= \frac{c}{3} \ln L + k \; ,
\end{equation}
where $c$ is the central charge of the associated conformal field theory, 
and $k$ is a non-universal constant, while $S_L$ was found to remain bounded 
in non-critical chains. This type of scaling of the entropy 
seemed to be related to the conformal invariance of homogeneous 
critical systems \cite{VLRK,CC}
. 
However, similar logarithmic scaling of the entropy was also found,
although, with a different "effective'' central charge, 
in critical inhomogeneous systems such as systems with 
a single defect \cite{Def1,Def2} or quenched random 
disorder \cite{RM,NL,CMCF,RS,HR,BY,saguia,refael2},
where the ground state and 
the low temperature properties are quite different from those of 
homogeneous models \cite{IM}.
Some of these recent results have already been used in 
the experimental identification of certain disordered phases \cite{rappaport}.

In the present paper a systematic study of the 
scaling of the mean entanglement entropy 
in a wide class of aperiodic models is carried out.
The considered models, the self-similarly 
singlet-renormalizable aperiodic XXZ chains,
are characterized by the property that
the aperiodic sequence of the couplings remain invariant under 
a real space RG procedure which produces effective spin singlets. 
The ground states of these models are aperiodic singlet
states, at least in the limit where the RG method applies.
Some features of two particular models from this class (the Fibonacci
and the tripling chain) have already been investigated in
a recent work \cite{IJZ}.

We shall derive a general formula for the effective central charge 
characterizing the asymptotic $L$-dependence of the average entropy
and show that in the XXZ chain this type of inhomogeneity may 
even enhance entanglement in contrast to single defect  
and random perturbations, in which cases the effective central 
charge decreases or remains invariant. 
Although, it is known that in some special chains of spins
with many components also random disorder may increase 
entanglement \cite{RS}.   
 
The paper is organized as follows.
In Sec.\ref{sec2} some basic features of aperiodic sequences and 
the renormalization group method for
inhomogeneous quantum chains are recapitulated and the class of models 
to be studied  is introduced. 
In Sec.\ref{sec3} we derive an analytical expression for
the entanglement entropy asymptotics in the considered 
aperiodic singlet phases, whereas in Sec.\ref{sec4} results of numerical 
investigations for the XX chain are
presented. Sec.\ref{sec5} is devoted to the discussion of the results. 
Finally, in the Appendix, we prove that among the self-similarly 
singlet-renormalizable aperiodic XXZ models
the one with strong tripling modulation has the largest effective
central charge.


\section{Aperiodic quantum spin chains}
\label{sec2}

We study the antiferromagnetic spin-${1\over 2}$ XXZ chain defined 
by the Hamiltonian
\be 
H=\sum_i J_i (S_i^xS_{i+1}^x+S_i^yS_{i+1}^y+\Delta_0 S_i^zS_{i+1}^z),
\ee
where the $S^{\alpha}_i$'s ($\alpha=x,y,z$) are spin-${1\over 2}$ operators,
the site-dependent couplings  $J_i>0$ are taken from a finite set
$\{J_a,J_b,\dots \}$ 
and are modulated according to 
some aperiodic sequence, whereas $\Delta_0$ ($0\le \Delta_0 \le 1$) is the 
anisotropy parameter which is initially site-independent, but, as we will
see later, it becomes site-dependent in the course of renormalization, as well.

\subsection{Aperiodic sequences}

The aperiodic sequences considered in this work are composed of
letters taken from a finite alphabet $\{a,b,c,\dots\}$ 
and are defined by an inflation rule which 
assigns to each letter a word (i.e. a finite sequence of letters)  
and the (infinitely many) repeated 
application of which generates the sequence. 
E.g. the well-known Fibonacci
sequence consists of two different letters, $a$ and $b$, and is 
defined by the inflation rule
\be 
\sigma_{F}: \left\{
\begin{array}{c}
 a \to w_a=ab \\
 b \to w_b=a.
\end{array}
\right. 
\label{fib}
\ee
The first few inflation steps starting from letter $a$ results in the
following subsequent strings:
$a,ab,aba,abaab,abaababa,\dots$.
Many properties of aperiodic sequences are encoded in the 
substitution matrix $M_{\alpha\beta}=n_{\alpha}(w_{\beta})$, where 
$n_{\alpha}(w_{\beta})$ denotes the number of letters $\alpha$ 
in the word $w_{\beta}$.
E.g. the largest eigenvalue $\lambda_+$ of $M$ gives the asymptotic 
ratio of lengths of the subsequent strings in the inflation procedure.

A finite aperiodic Hamiltonian is constructed by taking 
a finite section of the infinite aperiodic sequence and 
assigning a coupling $J_{\alpha}$ to link $(i,i+1)$ whenever 
the $i$th letter of the sequence is $\alpha$.

\subsection{Relevance/irrelevance criteria}

According to a heuristic criterion developed originally for random
models\cite{harris} and later generalized to aperiodic
systems\cite{luck}, the relevance of a weak aperiodic modulation is 
related to the wandering exponent $\omega$ which characterizes the 
fluctuations of the local order parameter. To be concrete,  
an aperiodic perturbation is predicted to be relevant if 
\be 
\omega > 1-\frac{1}{d\nu},
\label{hl}
\ee 
where $d$ is the dimension of the system and $\nu$ is the correlation
length exponent of the homogeneous system. 

For the particular case of aperiodic XX chains an exact relevance/irrelevance
criterion is available which is obtained by an exact renormalization
group method \cite{hermisson}. Here, the relevant wandering exponent 
is related to the fluctuations of the non-overlapping pairs of
letters, and the aperiodic perturbation is relevant (irrelevant) if 
$\omega >0$ ($\omega <0$), in agreement with the Harris-Luck criterion
with $d=1$ and $\nu=1$. For $\omega =0$, which is realized e.g. 
in the case of Fibonacci modulation, the perturbation is strictly marginal 
and critical exponents vary continuously with the coupling ratio.      

According to a weak perturbation renormalization group
study of the general XXZ chain \cite{vidal} the weak Fibonacci
modulation drives the system away from the pure system fixed point 
for any anisotropy $0\le \Delta_0\le 1$.

\subsection{Strong disorder renormalization group}

\begin{figure}[h]
\begin{center}
\includegraphics[width=0.6\linewidth]{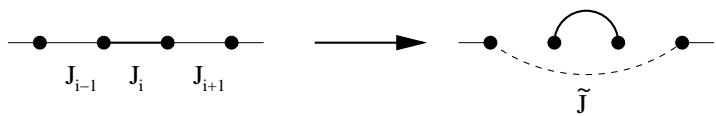}
\caption{\label{figsdrg} Renormalization scheme in the XXZ chain. The
  spins connected by the strongest bond $J_i$ form a singlet and a new
  bond $\tilde J$ is generated.} 
\end{center}
\end{figure}

Recently, many results have been obtained for aperiodic quantum spin
chains by means of a real space renormalization group method
\cite{hida,vieira} which was originally 
developed for random systems \cite{mdh,fisherxx}.
The main point of the method is the successive elimination of 
high-energy degrees of freedom connected to the largest 
coupling in the system. 
For the antiferromagnetic XXZ chain, the spin pair ($S_i,S_{i+1}$) 
coupled by the largest bond $J_i$, 
which is approximatively in the singlet state 
$\frac{1}{\sqrt{2}}(|\uparrow\downarrow\rangle-|\downarrow\uparrow\rangle )$
provided that $J_i\gg J_{i-1},J_{i+1}$,
$\Delta_iJ_i\gg \Delta_{i-1}J_{i-1},\Delta_{i+1}J_{i+1}$, is  
eliminated from the chain, and effective new couplings between 
spins $S_{i-1}$ and $S_{i+2}$  
\be 
\tilde J=\frac{J_{i-1}J_{i+1}}{(1+\Delta_i)J_i}, \qquad 
\tilde \Delta = \frac{1+\Delta_i}{2}\Delta_{i-1}\Delta_{i+1}
\label{rg}
\ee 
are generated by second order perturbation theory (see Fig. \ref{figsdrg}). 
This elimination step is iterated, leading to 
the gradual decrease of the number of active spins and that of the  
energy scale set by the actually largest effective coupling.
For the critical random XXZ chain the distribution of couplings is 
broadening without limits during the procedure, implying that the
conditions for the perturbation calculation are better and better
fulfilled, and it becomes asymptotically exact in
the (infinite randomness) fixed point. 
The latter is attractive for any amount of disorder, in
agreement with the Harris criterion which predicts the random
inhomogeneity to be relevant for $0\le \Delta_0\le 1$.  
The resulting ground state constructed by the method consists of
singlets of spins which are separated by arbitrarily large
distances, and the system is in the so called random singlet phase. 

\begin{figure}[h]
\begin{center}
\includegraphics[width=0.6\linewidth]{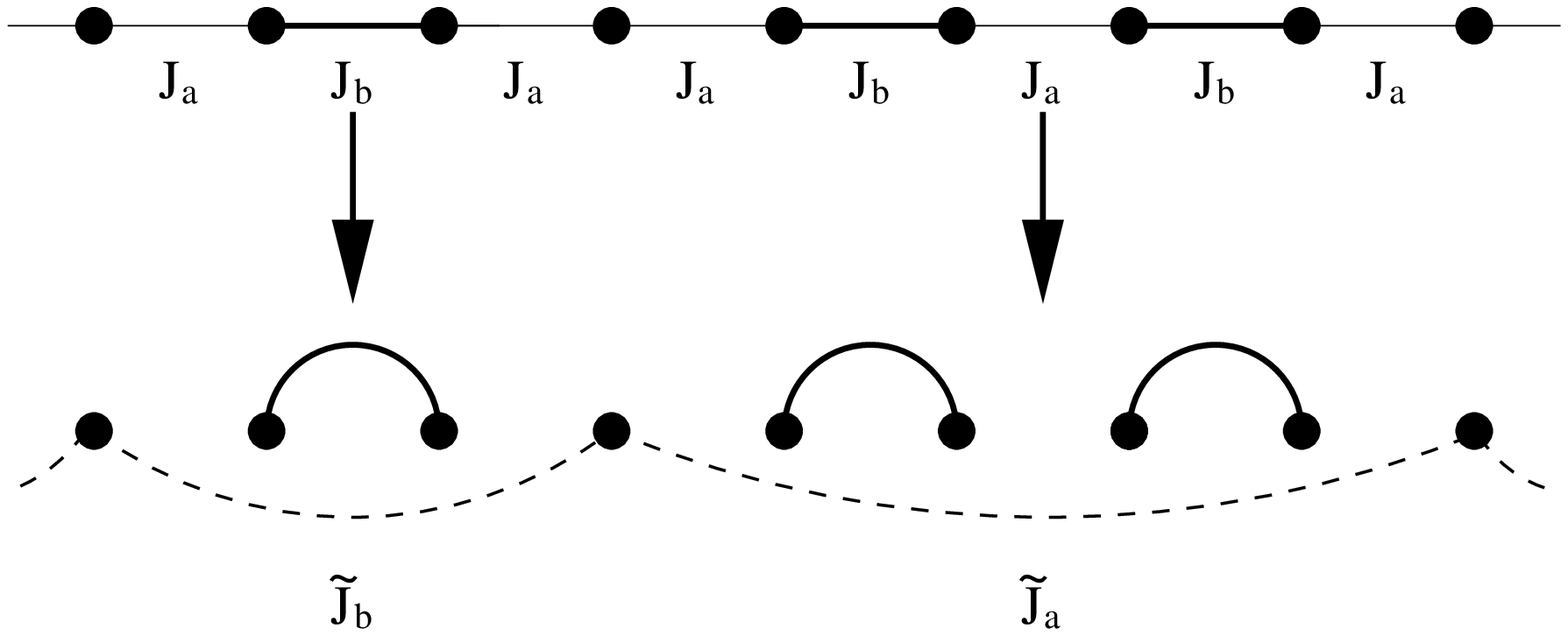}
\caption{\label{fibblock} Renormalization of the Fibonacci XXZ chain.} 
\end{center}
\end{figure}

The method was applied to the XXZ chain with various types of aperiodic 
modulations, where identical blocks are simultaneously renormalized
in the whole chain. After such a renormalization step the structure of
certain aperiodic modulations remain self-similar \cite{hida,vieira}. 
E.g. for the Fibonacci XXZ chain which is known to be such a system,
the blocks to be renormalized correspond to the words $aba$ and
$ababa$ (see Fig. \ref{fibblock}), and the renormalization step is essentially the reversed process
of the triple application of $\sigma_F$ given in eq. (\ref{fib}) \cite{hida}.
When $\Delta_0=1$, the ratio
of the couplings $r_{ab}\equiv J_a/J_b$ and the anisotropy renormalize
to $r_{ab}^*=0$ and $\Delta^*=1$, respectively. 
The method becomes asymptotically exact
in the fixed point which is attractive for any $r_{ab}<1$ 
in agreement with the fact that weak perturbation is relevant.
The ground state consists of spin singlets the possible lengths of 
which form a subset of the Fibonacci numbers and the system is 
in the so called {\it aperiodic singlet phase}. 
When $\Delta_0=0$, the ratio $r_{ab}$ remains 
invariant in the RG procedure, indicating that the Fibonacci
modulation is marginal.
Here, the decimation steps are exact 
only in the strong modulation limit $r_{ab}\to 0$. 
For the case $0< \Delta_0<1$ the coupling ratio and anisotropy
parameter renormalize to some $r_{ab}^*>0$ and $\Delta^*=0$, respectively. 
Thus, the strong perturbation is expected to be marginal whereas 
the weak perturbation is relevant. 
Here, the critical exponents are non-universal again, depending both on
$r_{ab}$ and $\Delta_0$ \cite{hida}.

\subsection{Singlet producing sequences}

In this work, we focus on 
XXZ chains with such aperiodic sequences 
(like the Fibonacci chain), 
where the RG procedure produces exclusively effective singlets 
and the sequence of couplings remains invariant.
Generally, in such systems, the blocks to be renormalized consist of an even
number of spins (an odd number of couplings), and the couplings in
each block are alternately weak and strong (with strong bonds at even
places) in order to avoid the forming of effective spins \cite{vieira}.
Furthermore, not only the RG-invariance of the 
sequence of couplings (self-similarity)
is required, but also the order of strength of the couplings 
should remain invariant under renormalization, i.e.  
$J_{\alpha}>J_{\beta}$ should imply $\tilde J_{\alpha}>\tilde J_{\beta}$.   
Each such sequence has a unique {\it canonical inflation rule} 
where the word $w_{\alpha}$ represents that block which just renormalizes 
to the two-spin block represented by letter $\alpha$, that is, 
the inflation rule
just corresponds to the reversed renormalization step.
To avoid some unphysical situations we also require two technical
conditions: the ergodicity of the inflation rule, i.e. the same infinite
sequence should be obtained in the bulk independently of the letter
which the iteration is started from, and the RG procedure should be ``unambiguous'',
i.e. $w_{\alpha} \ne w_{\beta}$ if $\alpha \ne \beta$. 
We call the sequences fulfilling the above conditions 
{\it singlet producing self-similar sequences}. 

The family of singlet producing self-similar
two-letter sequences have the following canonical inflation rule:
\be
\sigma_{mn}: \left\{
\begin{array}{c}
a \to w_a=aba(ba)^{m-1} \\
b \to w_b=a(ba)^{n-1},  
\end{array}
\right. 
\label{2letter} 
\ee
where the integers $m,n$ satisfy $1\le n\le m$, 
and letter $b$ now and in the rest of this work represents the
strongest bond $J_b$.  
The sequence generated by the transformation with $m=n=1$ is the
well-known silver-mean sequence, the choice $m=n=2$ generates the Fibonacci
sequence, whereas the transformation with $m=n=3$ can be
expressed as a composition of simpler inflation transformations 
as $\sigma_F\circ\sigma_{11}\circ\sigma_F$. 
An example for a singlet producing sequence composed of three
letters is 
\be
\sigma_{6-3}: \left\{
\begin{array}{c}
a \to w_a=abababc \\
b \to w_b=abc \quad \\
c \to w_c=ababc,   
\end{array}
\right. 
\label{6-3} 
\ee 
which is the sequence obtained after 
the first RG step of the 6-3 sequence \cite{vieira}. 
Another one is the tripling sequence
\be
\sigma_{t}: \left\{
\begin{array}{c}
a \to w_a=aba \\
b \to w_b=cbc \\
~c \to w_c=abc.   
\end{array}
\right. 
\label{tripling} 
\ee
which plays a special role among singlet producing sequences as will
be shown later. 
For the latter two sequences the relations $J_b>J_c>J_a$ remain invariant
under the renormalization.
Generally, the new effective couplings after an RG step are given in 
terms of the couplings of 
the previous generation and the canonical substitution matrix as
\beqn 
\tilde J_{\alpha}=
\frac{\prod_{\gamma\neq b}J_{\gamma}^{n_{\gamma}(w_{\alpha})}}{J_{b}^{n_{b}(w_{\alpha})}(1+\Delta_{b})^{n_{b}(w_{\alpha})}},
\nonumber \\
\tilde \Delta_{\alpha}=
\left(\frac{1+\Delta_{b}}{2}\right)^{n_{b}(w_{\alpha})}\prod_{\gamma\neq
  b}\Delta_{\gamma}^{n_{\gamma}(w_{\alpha})},
\eeqn  
which are obtained by eliminating the strong bonds $J_b$ one by one 
in the blocks represented by the words $w_{\alpha}$, using  
eq. (\ref{rg}). 
The considered sequences can be categorized into three types
on the basis of their relevance in the strong perturbation limit.

Those sequences 
for which the strong perturbation is relevant independently of
$\Delta_0$ are of type I. 
In this case, the coupling ratios
renormalize to zero, the procedure is asymptotically exact and the
resulting ground state is an aperiodic (independent) singlet state. 
These aperiodic modulations are exactly known to be relevant in the XX
model for arbitrary coupling ratios, and this is expected to hold also for
$0 < \Delta_0 \le 1$.   
The two-letter sequences with $m>n$ and e.g. the
6-3 sequence are of type I.
Type II sequences, such as the Fibonacci sequence, 
are relevant for the Heisenberg chain ($\Delta_0=1$),  
but are strictly marginal for the XX chain ($\Delta_0=0$). 
Here, for $\Delta_0=0$, the coupling
ratios remain invariant under the renormalization, whereas for 
intermediate initial anisotropies 
$0<\Delta_0<1$ the coupling ratios tend to a finite limiting value. 
So, the ground state is an aperiodic singlet state for $\Delta_0=1$,
but for $\Delta_0<1$ this is true only in the limit of vanishing initial
coupling ratios. The two-letter sequences with $m=n$ are of this
type.
Type III modulations are strictly marginal for the XX
model, and the coupling ratios remain invariant under renormalization,
independently of $\Delta_0$, thus marginality is indicated  
(at least in the strong modulation limit). 
Here, the ground state is an aperiodic singlet state only in the
limit of vanishing initial coupling ratios. The
tripling sequence and other sequences with words of equal length
belong to this class.  

\section{Entanglement entropy in the aperiodic singlet phase}
\label{sec3}

\subsection{Analytical expression for the scaling of entanglement entropy}

\begin{figure}[h]
\begin{center}
\includegraphics[width=0.6\linewidth]{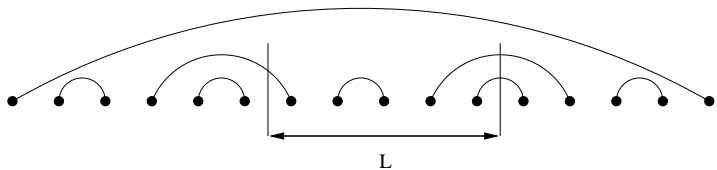}
\caption{\label{figblock} Calculation of the entanglement entropy. In
  the present case, the entanglement entropy of the block indicated by
  the arrow is $3\ln 2$.}  
\end{center}
\end{figure}

\begin{figure}[h]
\begin{center}
\includegraphics[width=0.6\linewidth]{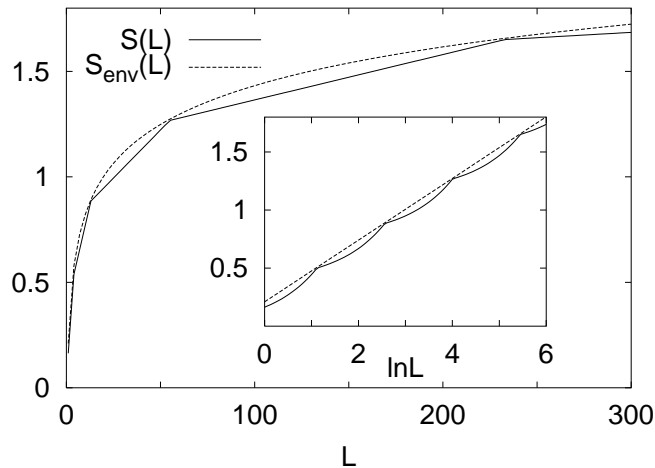}
\caption{\label{fig1} Average entanglement entropy as a function of the block
  size $L$ given in eq. (\ref{entropy}), evaluated for the Fibonacci sequence
  (solid curve). The dashed curve is the enveloping function 
    $S_{\rm env}(L)=\frac{c_{\rm eff}}{3}\ln L+{\rm const}$.} 
\end{center}
\end{figure}

In this section, we consider an infinite chain  
and calculate the average entanglement entropy of 
subsystems consisting of $L$ consecutive spins.
In the aperiodic singlet phase this can be done analytically, 
since, similarly to the random singlet 
phase, one simply has to count the singlet bonds connecting 
the subsystem with the rest of the system \cite{RM}. 
The entanglement entropy of one spin in the singlet pair is $\ln 2$,
therefore each such bond gives a contribution of $\ln 2$ to the 
entanglement entropy (Fig. \ref{figblock}).
At any stage of the renormalization procedure, the spin pairs, which 
form singlets, are those  connected 
by the actually strongest (and, at the same time, shortest) bonds, 
$\tilde J_b$. The evolution of bond lengths during the renormalization
can be 
determined by arranging them in a vector $w$ and noticing that the 
corresponding bond lengths after a renormalization step are obtained 
by multiplying $w$ by the transpose of the canonical substitution matrix $M$.
The length of the singlets in the $n$th generation (i.e. those generated in
the $n$th step of the renormalization), which is equal to the length of
the bond corresponding to letter $b$ before the $n$th step, 
can thus be calculated after diagonalizing $M$, yielding
\be 
l_n=u_b \lambda_+^{n-1} +
\mathcal{O} (\lambda_-^{n-1}),
\label{ln}
\ee  
where $\lambda_+$ is the largest eigenvalue of $M$ 
and $u_{\alpha}$  denotes the component $\alpha$ of the left eigenvector 
of $M$ associated with the eigenvalue $\lambda_+$.
This eigenvector and the right eigenvector $v$ 
of $M$ associated with $\lambda_+$  
are normalized as follows  
\beqn
\sum_{\alpha}v_{\alpha}=\sum_{\alpha}u_{\alpha} v_{\alpha} =1. \label{norm1}
\eeqn
The terms in the r.h.s. of eq. (\ref{ln}), denoted by $\mathcal{O}
(\lambda_-^{n-1})$, involve the subleading eigenvalues of $M$ and are 
negligible compared to the leading order term for large $n$.
The concentration of singlets of the $n$th generation, 
i.e. their average number per unit length, is given by 
\be 
\rho_n=v_b \lambda_+^{-(n-1)}.
\label{rn}
\ee  
The fraction of the chain which is covered by singlets of the $n$th
generation is $w_n=l_n\rho_n$, which tends rapidly to 
$u_b v_b$ for large $n$. This quantity gives the probability for
breaking a singlet bond of the $n$th generation when the chain is cut at 
a randomly chosen link.  
Therefore, each generation for which $l_n<L$ contributes a term of 
$2w_n\ln 2$ to the average entanglement entropy.  
The total contribution from singlets with $l_n<L$ is thus 
for large $L$ 
\be
S_1(L)\approx 2n(L)u_b v_b \ln 2+{\rm const},
\ee
where $n(L)$ is the number of generations with $l_n<L$.
From eq. (\ref{ln}) we get for large $L$   
\be
n(L)=\left[\frac{\ln (L/u_b)}{\ln \lambda_+}\right]_{\rm int},
\label{n}
\ee
where $[x]_{int}$ stands for the integer part of $x$.
The contribution of singlets with $l_n>L$ is 
\be
S_2(L)= \sum_{n=n(L)+1}^{\infty}2\rho_nL\ln 2= L\lambda_+^{-n(L)-1}\ln 2,
\ee
where we have used the relation 
\be 
\sum_{n=1}^{\infty}\rho_n=\frac{1}{2},
\label{sumrule}
\ee
which expresses the fact that all spins are paired in the aperiodic
singlet phase.    
Finally, we obtain for the asymptotic $L$-dependence of the 
entanglement entropy:
\be
S(L)= 2n(L)u_b v_b \ln 2+L\lambda_+^{-n(L)-1}\ln 2
+{\rm const}.  
\label{entropy}
\ee
As opposed to translation invariant systems, $S(L)$ is 
a piecewise linear function with breaking points at all
possible singlet lengths $l_n$, see Fig. \ref{fig1}.
The appearance of the characteristic
lengths $l_n$ in the correlation functions is common in 
the aperiodic singlet phase \cite{vieira}.   
Nevertheless, the enveloping curve of $S(L)$, which asymptotically
fits to the breaking points, grows logarithmically with 
$L$, hence one can write the entropy in the form
\be
S(L)=\frac{c_{\rm eff}}{3}\ln L+P(L),
\label{envelope}
\ee
where $P(L)$ is an oscillating log-periodic function. 
From eqs. (\ref{n}) and (\ref{entropy}) we obtain the general 
expression for the effective central charge 
\be 
c_{\rm eff}=6\ln 2 \frac{u_b v_b}{\ln \lambda_+},
\label{ceff}
\ee
which differs
from the central charge  of the homogeneous system $c=1$ and 
it is characteristic for the underlying aperiodic sequence.
Evaluating eq. (\ref{ceff}) for the two-letter sequences $\sigma_{mn}$ 
yields 
\be 
c_{\rm eff}=
\frac{6\ln 2}{\ln
  \lambda_+}\frac{(\lambda_+-m-1)^2}{mn+(\lambda_+-m-1)^2},
\ee
with the rescaling factor 
$\lambda_+=\frac{1}{2}(m+n+\sqrt{(m+n)^2+4(m-n+1)})$.
For the 6-3 sequence we get 
$c_{\rm eff}=\frac{36}{25}\frac{\ln 2}{\ln5}\approx 0.6201$, whereas for
the tripling sequence $c_{\rm eff}=\frac{\ln 4}{\ln 3}\approx 1.2618$.
\begin{table}[h]
\begin{center}
\begin{tabular}{|l|r|}
\hline m=n=1 (silver mean) & 0.6910... \\
\hline m=n=2 (Fibonacci) & 0.7962... \\
\hline m=n=3  & 0.7819... \\
\hline m=n=4 & 0.7519... \\
\hline m=n=5 & 0.7228... \\
\hline
\end{tabular}
\end{center}
\caption{\label{tab} Effective central charges for aperiodic sequences defined 
in eq. (\ref{2letter}) in the aperiodic singlet phase.} 
\end{table}
 
\subsection{The maximally entangled aperiodic singlet topology}

Next, we turn to discuss the possible bounds of the effective
central charge. Considering the two-letter sequences, $\sigma_{mn}$, 
for the sake of simplicity with $m=n$, we have found that after 
the maximum at the Fibonacci sequence ($m=n=2$) the effective central charge 
is decreasing with the length of the words (see Table I.) 
and it finally tends to zero in the limit $n=m\to\infty$. 
This is easy to understand, since the words themselves
are periodic, and the blocks corresponding to the words have a
dimerized (non-critical) ground state. Thus, the effective central charge
can be arbitrarily close to zero. One may ask whether 
sequences (belonging to the considered class)
with an arbitrarily large effective central charge can be constructed. The answer is negative. 
To see this at a heuristic level, note that the expression 
for the effective central charge in (\ref{ceff})
may be separated into two (non-independent) factors, 
the covered fraction and the rescaling
factor. When the singlets of the same generation lie
densely, leaving only a small concentration of unpaired spins between
them, the covered fraction will be large which is favorable for 
strong entanglement, however, the rescaling factor will also be large,
which has the opposite effect. Conversely, for a large concentration of lonely
spins, the rescaling factor is moderate, however, the covered fraction is
also small. Thus, one expects that there may be an optimal aperiodic 
singlet topology which maximizes the entanglement entropy, and indeed
this is the case. We can prove that the aperiodic state belonging
to the tripling sequence has the highest central charge (see the Appendix).
Beside this proof, one can also give a less formal, but more suggestive
argument for this. 
\begin{figure}[t]
\begin{center}
\includegraphics[width=0.6\linewidth]{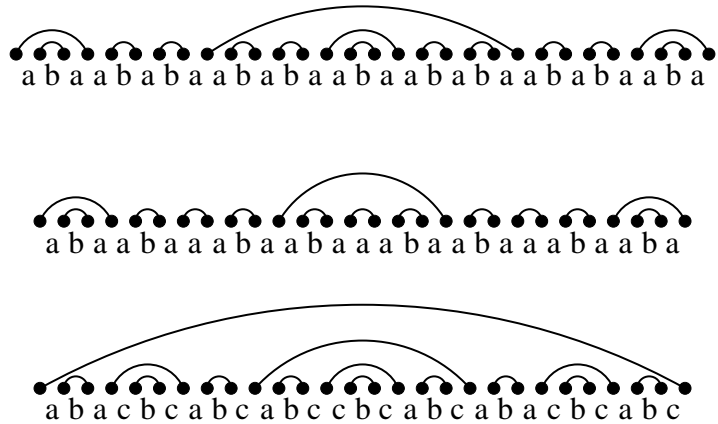}
\caption{\label{singtop} The singlet bond topology of the Fibonacci
(upper figure),  the silver-mean (middle figure), and the 
tripling (lower figure) sequences.}  
\end{center}
\end{figure}
The singlet arrangement 
produced by the tripling sequence (see Fig. \ref{singtop}) 
is optimal in the following 
sense:  Regarding those subsequent renormalization steps which generate 
non-overlapping singlets (e.g. for the silver mean sequence pairs of subsequent
steps), as one single renormalization step (in which singlets of 
different lengths may be produced), the number of unpaired spins between 
singlets of the same generation is at most one. For the tripling sequence 
there is an unpaired spin between every neighboring singlet, 
which is the optimal situation. 

\section{Numerical analysis}   
\label{sec4}

In order to check the results obtained in the previous section 
and to get some insight into the behavior of
entanglement entropy for marginal perturbations when the ground state
is not an independent singlet state and the RG breaks
down, we performed numerical calculations for the XX model ($\Delta_0=0$). 
This model can be mapped to a system of free fermions \cite{lsm} 
and the reduced
density matrix of the subsystem can be written as the exponential of a
free-fermion operator \cite{peschel}.  The entanglement 
entropy can be computed from the eigenvalues $\lambda_i$ of the 
correlation matrix of fermion 
operators $C_{ij}=\langle c_i^{\dagger}c_j\rangle$ restricted to the
subsystem by the formula \cite{VLRK,peschel}
\be 
S(L)=-\sum_i\left[\lambda_i\ln \lambda_i+ 
(1-\lambda_i)\ln(1-\lambda_i)\right].
\label{ent}
\ee   
The computation of the correlation matrix necessitates the
diagonalization of an $N\times N$ matrix, then diagonalizing the
restricted correlation matrix of size $L\times L$ one obtains the von
Neumann entropy from eq. (\ref{ent}).  

We performed numerical calculations for chains with both free
and periodic boundary conditions, 
and calculated the von Neumann entropy of blocks which were located in
the former case in the middle of the chain.
The size of blocks varied from $2$ to $N/2$ and 
the largest system size we considered was $N=1024$.
The entropies were then averaged over a few thousand independent samples, 
whereas for periodic chains an averaging over the different starting 
positions of the block was also carried out.
 
\begin{figure}[h]
\begin{center}
\includegraphics[width=0.6\linewidth]{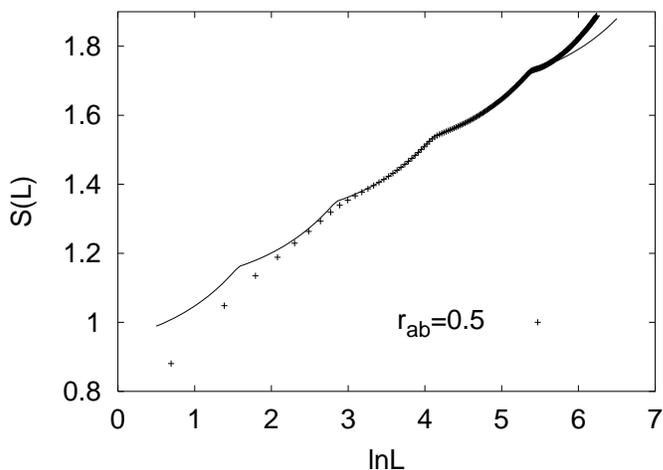}
\caption{\label{figrel} Numerically calculated average 
entanglement entropy as a function of the block
  size $L$ for the relevant two-letter sequence defined in
  (\ref{2letter}) with $m=2$ and $n=1$. The calculations were
  performed with free boundary condition. The size of the system is
  $N=1024$ and the data were averaged over $10^4$ samples. The
  solid line is the analytical prediction (\ref{entropy}). For this
  sequence $c_{\rm eff}\approx 0.4459$.}
\end{center}
\end{figure}
\begin{figure}[h]
\begin{center}
\includegraphics[width=0.6\linewidth]{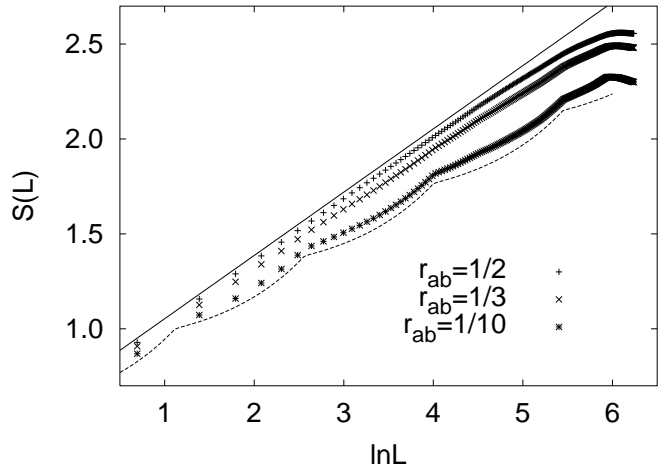}
\caption{\label{figfib} Numerically calculated average 
entanglement entropy as a function of the block
  size $L$ in the Fibonacci XX chain, for different coupling ratios. 
  The calculation was performed with
  periodic boundary condition. The size of the system is $N=1024$ and
  the data were averaged over $20$ independent samples and over the
  different starting positions of the blocks. The solid line is the
  asymptotics of $S(L)$ for the homogeneous case $r_{ab}=1$, whereas the
  dashed curve is the expected asymptotical curve in the limit
  $r_{ab}\to0$, given in eq. (\ref{entropy}).}    
\end{center}
\end{figure}
The simplest sequence among the two-letter sequences $\sigma_{mn}$
which is of type I and is thus relevant
even in the XX limit is the sequence with $m=2$ and $n=1$. The average
entropy as a function of the block size, shown in
Fig. \ref{figrel}, tends to the asymptotic curve characteristic for
the independent singlet phase after a crossover region even for a
finite initial coupling ratio in
accordance with the expectations.  
\begin{figure}[h]
\begin{center}
\includegraphics[width=0.6\linewidth]{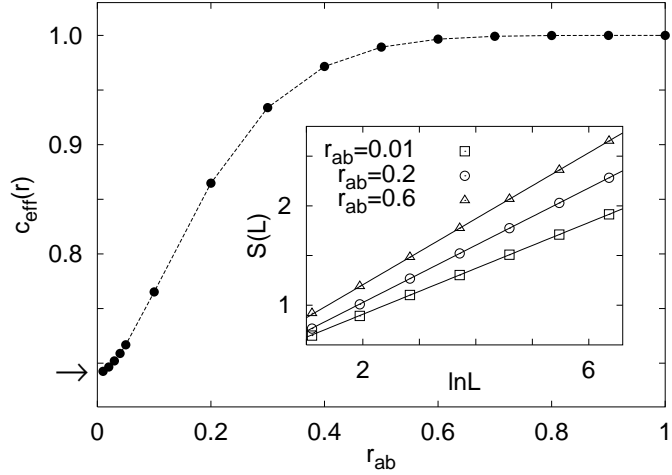}
\caption{\label{figsilver} Effective central charge for different 
coupling ratios $r_{ab}$ calculated for the silver-mean XX chain with
periodic boundary condition. The arrow indicates the exactly
calculated value in the aperiodic singlet phase, taken from Table \ref{tab}. 
Inset: The average entropy for different
block sizes $L=N/2$ which are chosen to be the first few singlet bond
lengths $l_n$ in order to avoid log-periodic oscillations. 
The averaging was performed on the $L$ different starting
positions of the block.}    
\end{center}
\end{figure}

\begin{figure}[h]
\begin{center}
\includegraphics[width=0.6\linewidth]{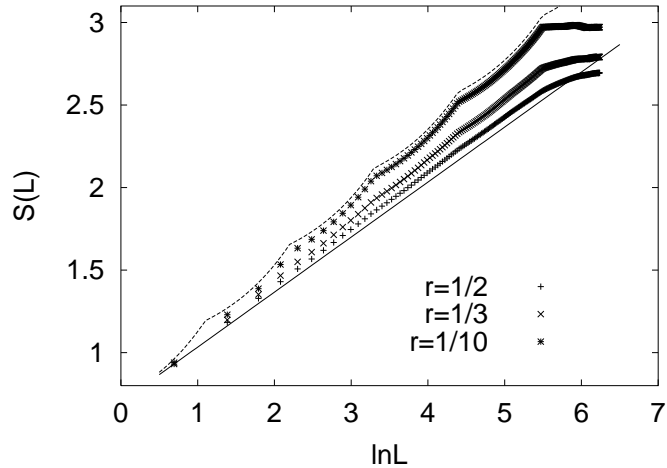}
\caption{\label{figrt} Numerically calculated entanglement entropy as a function of the block  size $L$ in the tripling XX chain, for different values of the
  coupling ratios $J_a/J_c=J_c/J_b\equiv r$. 
 The calculations were performed with free boundary condition.
 The size of the system is  $N=1024$ and the data were averaged over $10^4$ samples. The solid line is the asymptotics of $S(L)$ for the homogeneous case $r=1$, whereas the
 dashed curve is the expected asymptotical curve in the limit $r\to 0$, given in (\ref{entropy}).} 
\end{center}
\end{figure}

For type II sequences, which are marginal
in the XX limit and relevant in the Heisenberg limit, one expects that,
analogously to the critical exponents, 
the effective central charge varies with the coupling ratio
 and the anisotropy parameter for $\Delta_0<1$.  
Numerical results in the XX case for the size dependence of the entanglement 
entropy for one representant of this class, the Fibonacci
sequence, are shown in Fig. \ref{figfib} for different coupling 
ratios $r_{ab}$. 
As expected, the effective central charge,
which is related to the slope of the straight line fitted to the breaking
points of the curves in the semi-logarithmic plot, varies
continuously with the coupling ratio, see 
also the numerical results in \cite{IJZ}. 
For an other type II sequence, the silver-mean sequence, the effective
central charge as a function of the coupling ratio is shown in 
Fig. \ref{figsilver}.

Type III sequences, being marginal for any $\Delta_0$, are
also expected to have effective central charges which may depend on
the coupling ratios and the anisotropy. Numerical results for the
tripling sequence in the XX limit, presented in Fig. \ref{figrt}, are in
agreement with these expectations. 

\section{Discussion}   
\label{sec5}

We have studied in this work the average entanglement entropy of blocks of
contiguous spins in a class of aperiodic quantum spin chains, where
the sequence of couplings remains invariant under the strong disorder RG
procedure which produces effective spin singlets. 
For these systems, the ground state is, in the limit where
the RG method applies, an aperiodic (independent) singlet state.
Knowing the topology of singlet bonds,
the average von Neumann entropy as a function of the block size can be
calculated analytically. This quantity, apart from the log-periodic 
oscillations, is found to grow logarithmically with the size of the block.
A general expression for the effective central charge was derived,
and the modulation producing the largest central charge was found.

Beside the singlet producing self-similar sequences studied in this work, 
which can be termed as "pure", also other singlet 
producing sequences can be constructed by composing 
canonical inflation transformations of different pure sequences. 
An example is the transformation 
$\sigma_{11}\circ\sigma_{33}$ which is, on the other hand, 
just the square of the transformation
$\sigma_{FS}\equiv\sigma_{11}\circ\sigma_F$, therefore both generate
the same "golden-silver-mean" sequence. 
For composite sequences involving $m$ consecutive pure 
transformations,
$M_1,M_2,\dots,M_m$, the calculations outlined in Sec. \ref{sec3} can be 
generalized in a straightforward way. In this case the resulting effective 
central charge is the sum of central charges calculated from the $m$
cyclic permutations of the product $M_1\circ M_2\circ\dots\circ M_m$ using 
eq. (\ref{ceff}). 
For example, for the golden-silver-mean sequence we obtain 
$c_{\rm eff}=\frac{6(\sqrt{3}-1)}{\sqrt{3}\ln (7+4\sqrt{3})}\ln
2\approx 0.6673$. 

Finally, we mention that for sequences of type I and II, 
which are relevant at least for $\Delta_0=1$, where 
the perturbation of arbitrary strength is expected to drive the system 
to the strongly aperiodic fixed point, 
we have not found larger values of $c_{\rm eff}$ than that of 
the homogeneous model $c=1$. 
This result is consistent with the "generalized c-theorem"
proposed in Ref \cite{RM} as a possible
extension of the original c-theorem \cite{ctheorem} 
to disordered systems. 
According to this conjecture, the effective central charge,
defined through entanglement scaling, decreases
both for homogeneous and disordered systems
along RG trajectories induced by relevant perturbations. 
The sequences for which 
we have found $c_{\rm eff}>1$, 
the tripling and a sequence with words of length five, 
both belong to type III. 
These types of perturbations are 
marginal at least at the strongly aperiodic fixed point (in the
special case $\Delta_0=0$ even for perturbations of arbitrary strength),  
therefore these cases are out of the range of
validity of the generalized c-theorem. 
However, for $\Delta_0>0$, the behavior of the effective central charge for finite perturbations of this type is still an open question. 
It is worth mentioning that our results show that
disordered marginal perturbations can
increase the effective central charge,
which is not expected in case of homogeneous marginal
perturbations \cite{cardy}.

\ack
We thank F. Igl\'oi for useful discussions. 
This work has been supported by the National Office of Research and 
Technology under grant No. ASEP1111, by the Hungarian National
Research Fund under grant No. T043149 and by the Deutsche
Forschungsgemeinschaft under grant No. SA864/2-2. 

\appendix 
\section{}

Here, we prove that the tripling state has the largest 
effective central charge among the states generated by singlet producing 
self-similar sequences. 

First, we notice that the relations (\ref{rn}) and (\ref{sumrule}) 
imply $v_b=\frac{1}{2}(1-\lambda^{-1}_+)$. Inserting this 
into the expression of the central charge in eq. (\ref{ceff}) we obtain
\be 
c_{\rm eff}= 3\ln 2 \frac{(1-\lambda^{-1}_+)u_{b}}{\ln \lambda_+}.
\label{effc2}
\ee
In the considered class of models, as we have mentioned,  
the strongest effective bond is at the same time the shortest one 
(or, more precisely, its length is smaller than or equal to that of the 
other effective bonds) after each renormalization step. 
Therefore, the length $l_{n+1}$ of singlets produced in the $n+1$st 
renormalization step cannot exceed the average distance  $\xi_n$ 
between active (unpaired) spins after the $n$th step, 
i.e. $\xi_n \ge l_{n+1}$. 
The concentration of active spins decreases exactly by a factor 
of $\lambda_+$ in each step, therefore $\xi_n=\lambda^n$, 
which, together with eq. (\ref{ln}) leads to $u_b \le 1$.
Putting this in eq. (\ref{effc2}) yields
\be 
c_{\rm eff}\le 3\ln 2 \frac{1-\lambda^{-1}_+}{\ln \lambda_+}.
\label{ineq}
\ee
Note that we have immediately an upper bound for all the
considered sequences, since the r.h.s. of (\ref{ineq}) is monotonously 
decreasing with $\lambda_+$($>1$), hence  $c_{\rm eff}< 3\ln 2$.
However, this bound can be further sharpened, as we show below.

For  $\lambda_+ \ge 3$, the statement we want to prove follows
directly, since in this case the maximal value of 
the upper bound (\ref{ineq}) is $\frac{\ln 4}{\ln 3}$, 
which is just the effective central
charge of the tripling state.

If $\lambda_+ < 3$, then letter $b$ which represents the strongest bond 
must be transformed by the 
canonical inflation rule into a single-letter word.
Indeed, $w_b=abc \dots$ would imply 
$l^b_{n+1} = l^a_n+ l^b_n + l^c_n + \dots \ge 3 l^b_{n}$, 
which is equivalent to  $\lambda_+ \ge  3$ .
Here, $l_n^{\alpha}=u_{\alpha}\lambda_+^{n}+\mathcal{O} (\lambda_-^{n})$ 
denotes the length of the bond represented by letter $\alpha$ 
(or shortly, bond $\alpha$)
after the $n$th renormalization step. 
The most general form of canonical inflation rules with $\lambda_+<3$, is thus
\beqn
\left\{
\begin{array}{ccc}
a_{1}^{(1)}  & \to  &a_{2}^{(1)}  \nonumber \\
a_{2}^{(1)} & \to  &a_{3}^{(1)}  \nonumber \\
  & \vdots &  \nonumber \\
a_{n_1-1}^{(1)} & \to  &a_{n_1}^{(1)}  \nonumber \\
a_{n_1}^{(1)} & \to  & w^{(1)}  \nonumber \\
\end{array}
\right. \\
\left\{
\begin{array}{ccc}
a_{1}^{(2)} & \to  &a_{2}^{(2)}  \nonumber \\
a_{2}^{(2)} & \to  &a_{3}^{(2)}  \nonumber \\
 & \vdots &  \nonumber \\
a_{n_2 -1}^{(2)} & \to  &a_{n_2}^{(2)}  \label{subs} \\
a_{n_2}^{(2)} & \to  & w^{(2)}    \nonumber \\
\end{array}
\right.  \\
 \qquad \quad \quad ~~\vdots &  &  \nonumber \\
\left\{
\begin{array}{ccc}
a_{1}^{(k)} & \to  &a_{2}^{(k)}  \nonumber \\
a_{2}^{(k)} & \to  &a_{3}^{(k)}  \nonumber \\
   & \vdots &  \nonumber \\
a_{n_k -1}^{(k)} & \to  &a_{n_k}^{(k)}  \nonumber \\
a_{n_k}^{(k)} & \to  & w^{(k)}  
\end{array}
\right.
\eeqn
where $a_{i}^{(j)}$ denotes a single letter, specially 
$a_{1}^{(1)}\equiv b$ is the letter representing the strongest bond, whereas 
$w^{(j)}$ stands for a word comprising at least three letters.
The indices $n_i$ are arbitrary positive integers with the restriction 
$n_1>1$.
In the following, we use the notation $v_{i}^{(j)}$ for the density of letter 
$a_{i}^{(j)}$ and write $u_{i}^{(j)} \lambda_{+}^n$ for the asymptotical length of the corresponding bond after the $n$th renormalization step. 
With these notations, eq. (\ref{norm1}) reads
\be
\sum_{j=1}^{k} \sum_{i=1}^{n_j} v_{i}^{(j)}u_{i}^{(j)}=1.
\label{norm2}
\ee
Furthermore, we need the following relations 
\beqn
v_{i+1}^{(j)}& \ge &v_{i}^{(j)}/\lambda_{+},  \label{cons1} \\
u_{i+1}^{(j)}& = & \lambda_{+} u_{i}^{(j)}, \label{cons2}\\
\lambda_{+} & \ge & \sqrt[n_1]{3},  \label{cons3} 
\eeqn 
the first two of which follow simply from the form of the inflation
rules. Applying eq. (\ref{cons2}) $n_1-1$ times for 
$j=1$ yields 
\be
u_{n_1}^{(1)}=\lambda_{+}^{n_1-1} u_{1}^{(1)}.
\label{chain} 
\ee
Using that $a_{n_1}^{(1)}$ is transformed to the long word $w^{(1)}$ 
and that $u_1^{(1)}\le u_i^{(j)}$ for all $i,j$, 
we have $\lambda_{+} u_{n_1}^{(1)} \ge 3u_{1}^{(1)}$. This inequality
together with eq. (\ref{chain}) give immediately relation (\ref{cons3}).

Now, we consider the group of letters $a_{i}^{(j)}$ with a fixed index
$(j)$ and introduce the quantities 
$v^{(j)}=\sum_{i=1}^{n_k} v_{i}^{(j)}$ and 
$w_i^{(j)}=v_{i}^{(j)}/v^{(j)}$, such that $\sum_{i=1}^{n_j}
w_{i}^{(j)}=1$ holds and  relations (\ref{cons2}) imply
\be
 1 \le \frac{w_{2}^{(j)}}{w_{1}^{(j)}}\lambda_{+} \le 
\frac{w_{3}^{(j)}}{w_{1}^{(j)}} \lambda_{+}^{2} \le \dots \le \frac{w_{n_j}^{(j)}}{w_{1}^{(j)}} \lambda_{+}^{n_j-1}.
\label{w}
\ee
With these variables we can write    
\be
\sum_{i=1}^{n_j} v_{i}^{(j)} u_{i}^{(j)}= 
u_{1}^{(j)} \sum_{i=1}^{n_j}v_{i}^{(j)} \lambda_{+}^{i-1}=
u_{1}^{(j)}v^{(j)} \sum_{i=1}^{n_j}w_{i}^{(j)} \lambda_{+}^{i-1},
\label{sum1}
\ee
where the first equality follows from eq. (\ref{cons2}).
The latter expression can be written also in the form 
\be
\sum_{i=1}^{n_j} v_{i}^{(j)} u_{i}^{(j)}= 
u_{1}^{(j)}v^{(j)}
\frac{ 1 +\frac{w_{2}^{(j)}}{w_{1}^{(j)}}\lambda_{+} + 
\frac{w_{3}^{(j)}}{w_{1}^{(j)}} \lambda_{+}^{2} + \dots +
\frac{w_{n_j}^{(j)}}{w_{1}^{(j)}} \lambda_{+}^{n_j-1}}{ 1 +
  \frac{w_{2}^{(j)}}{w_{1}^{(j)}} + 
\frac{w_{3}^{(j)}}{w_{1}^{(j)}}+ \dots +
\frac{w_{n_j}^{(j)}}{w_{1}^{(j)}}}.
\label{long}
\ee
One can show that, since $\lambda_{+}>1$, a lower bound of this 
 expression is obtained by substituting the relations
(\ref{w}) as equalities into the fraction on the r.h.s. of
 eq. (\ref{long}), and we get
\be
\sum_{i=1}^{n_j} v_{i}^{(j)} u_{i}^{(j)} \ge 
u_{1}^{(j)} v^{(j)} \frac {n_j\lambda_{+}^{n_j-1}(\lambda_{+}-1)}{\lambda_{+}^{n_j}-1}.
\ee
We know that  $u_{1}^{(j)} \ge u_{1}^{(1)}$, however, we can sharpen this bound if
$n_j < m$, where $m$ is the positive integer for which 
$\sqrt[m]{3} \le \lambda_{+} < \sqrt[m-1]{3}$.
(From (\ref{cons3}) it follows that $m \ge n_1>1$.) In this case,
(\ref{cons2}) yields $u_{1}^{(j)} \ge 3u_{1}^{(1)}/\lambda_{+}^{n_j}$.  
Using these relations we obtain
\be
\sum_{i=1}^{n_j} v_{i}^{(j)} u_{i}^{(j)} \ge  u_{1}^{(1)} v^{(j)} P_j(\lambda_{+}),
\ee
with
\be
P_j(\lambda_{+}) = \left\{
\begin{array}{c}
\frac {n_j\lambda_{+}^{n_j-1}(\lambda_{+}-1)}{\lambda_{+}^{n_j}-1} 
\; {\rm if } \; n_j \ge m \\
\frac {n_j3(\lambda_{+}-1)}{\lambda_+(\lambda_{+}^{n_j}-1)} 
\; \; {\rm if } \; n_j < m.
\end{array}
\right. 
\label{a5} 
\ee 
Summing over the index $(j)$ and using the normalization (\ref{norm2}) 
we get 
\be
1\ge u_{1}^{(1)}\sum_{j=1}^{k} v^{(j)} P_j(\lambda_{+}) \ge  u_{1}^{(1)} \min\{P_j(\lambda_{+})\},
\label{min}
\ee
where the latter inequality follows from $\sum_{j=1}^{k} v^{(j)}=1$.
It is easy to check that in the interval $\sqrt[m]{3} \le
\lambda_+ < \sqrt[m-1]{3}$
\be
 \frac{(1-\lambda^{-1}_+)}{P_j(\lambda_+)\ln \lambda_+} 
\le \frac{2}{3\ln{3}}
\ee
holds for all $P_j(\lambda_+)$. From this inequality and (\ref{min}) 
we obtain an upper bound on $ u_{1}^{(1)}$
\be
u_b\equiv u_1^{(1)} \le \frac{2 \ln \lambda_+}{3 \ln{3(1-\lambda^{-1}_+)}}.
\ee
Substituting this into the formula of the
effective central charge (\ref{effc2})  yields
\be
c_{\rm eff}\le \frac{\ln 4}{\ln 3}.
\ee
The obtained upper bound is again just the value of the 
effective central charge of the tripling state. 
Thus, we have completed the proof.



\end{document}